\documentclass[a4paper,11pt]{article}
\usepackage{pos}
\setcitestyle{maxnames=1}

\usepackage{subcaption, graphicx}
\usepackage{graphicx}

\title{Evidence of current-enhanced excited states in lattice QCD three-point functions}

\author*[a]{Lorenzo Barca}

\affiliation[a]{John von Neumann-Institut für Computing NIC,\\
  Deutsches Elektronen-Synchrotron DESY, \\
  Platanenallee 6, 15738 Zeuthen, Germany}

\emailAdd{lorenzo.barca@desy.de}

\abstract{
Excited-state contamination remains one of the leading sources of systematic uncertainty in the precise determination of hadron structure observables from lattice QCD. In this work, I present a general mechanism, motivated by meson dominance and implemented through the variational method, that identifies which excited states are enhanced by the choice of inserted current and kinematics. The argument is supported by numerical evidence and predictions from chiral perturbation theory across different hadronic channels, in particular in the nucleon sector, and provides both conceptual insight and practical guidance for controlling excited-state effects in hadron three-point function analyses.
}

\FullConference{The 42nd International Symposium on Lattice Field Theory (LATTICE2025)\\
2-8 November 2025\\
Tata Institute of Fundamental Research, Mumbai, India\\}

\begin{document}
\begin{flushright}
    \texttt{DESY-26-020}
\end{flushright}
\maketitle

\section{Introduction}

Hadron matrix elements of local currents, $\langle H'|J|H\rangle$, provide the nonperturbative input required to connect experimental observables to QCD. They enter a wide range of phenomenological applications, including precision studies of weak interactions in neutrino-nucleus scattering, dark-matter searches, and flavor physics. Reliable first-principles determinations from lattice QCD are therefore essential.

A major source of systematic uncertainty in these calculations is excited-state contamination (ESC): contributions from higher-energy states that bias the extraction of ground-state matrix elements from Euclidean three-point functions. While excited states are exponentially suppressed at large source-sink separations, practical calculations are limited to moderate separations due to the exponential degradation of the signal-to-noise ratio. Consequently, ESC remains large in most channels.

Various strategies have been developed to control ESC. Analysis methods such as multi-state fits to two- and three-point functions and the summation method aim to account for ESC explicitly. In practice, such analyses are often combined with the use of extended interpolating operators, designed to enhance the overlap with the ground state and reduce contributions from higher excited states.
Nevertheless, sizable excited-state effects persist in specific channels. In the nucleon sector, it has long been observed that even source-sink separations of $\approx 1.7~\mathrm{fm}$ do not guarantee negligible contamination.

Chiral perturbation theory (ChPT) provides a physical interpretation of this behavior by identifying low-lying multi-hadron states, such as $N\pi$ states, as dominant contributors in certain channels of the nucleon three-point functions. Incorporating this structure into multi-state fit strategies has led to improved consistency with symmetry relations such as the generalised Goldberger–Treiman relation, demonstrating that ESC can qualitatively affect extracted matrix elements.

The central message of this work is that large ESC can be systematically understood from symmetry and finite-volume enhancement arguments \cite{Barca:2025det}. While the numerical evidence is presented in the nucleon channel, the underlying mechanism is general and therefore applies to hadronic three-point functions more broadly. By identifying the origin of these effects, we provide guidance for the development of analysis strategies and operator constructions aimed at systematically controlling or removing ESC in precision lattice QCD calculations.
 \section{Extracting hadron matrix elements from three-point functions}

Hadron matrix elements are extracted from Euclidean three-point correlation functions
\begin{align}
C_{\rm 3pt}(\vec{p}', t; \vec{q}, \tau)
= 
\langle \mathrm{O}_{H_f}(\vec{p}', t)\,
\mathcal{J}(\vec{q}, \tau)\,
\bar{\mathrm{O}}_{H_i}(\vec{p}, 0) \rangle ,
\end{align}
whose spectral decomposition reads
\begin{align}
C_{\rm 3pt}(\vec{p}', t; \vec{q}, \tau)
=
\sum_{n,m}
Z^{|n(\vec{p}')\rangle}_{\mathrm{O}_{H_f}}\,
\bar Z^{|m(\vec{p})\rangle}_{\mathrm{O}_{H_i}}\,
\langle n(\vec{p}')|\mathcal{J}(\vec{q})|m(\vec{p})\rangle\,
e^{-E'_n(t-\tau)} e^{-E_m \tau}.
\end{align}
The overlap factors $Z$ and $\bar{Z}$ encode the coupling of the interpolating operators to the states with the same quantum numbers. In finite volume, their normalization depends on the number of particles in the state.
The same overlaps appear in the two-point function
\begin{equation}
C_{\rm 2pt}(\vec{p}, t)
=
\langle \mathrm{O}_{H_i}(\vec{p}, t)\,
\bar{\mathrm{O}}_{H_i}(\vec{p}, 0) \rangle
=
\sum_n |Z^{|n(\vec{p})\rangle}_{\mathrm{O}_{H_i}}|^2 ~e^{-E_n t}.
\end{equation}
In principle, ground-state matrix elements can be extracted from three-point functions at large source-sink separations, where contributions from excited states are exponentially suppressed. However, due to the signal-to-noise deterioration \cite{Parisi:1983ae}, the analysis is restricted to moderate source-sink separations, where ESC is not negligible. 
Multi-level sampling algorithms are very promising alternatives for reaching much larger source-sink separations \cite{Barca:2025dca, Barca:2024fpc, Barca:2023arw}

In practice, $\langle H_f | \mathcal{J} | H_i\rangle$ are usually  extracted either from simultaneous multi-state fits to two- and three-point functions or to suitable ratios in which the leading overlap factors cancel and the contribution from the ground state is $t$- and $\tau$-independent. 
In the latter case, any residual time dependence is a clear sign of ESC.
Recently, novel methods based on the Lanczos algorithm have been developed for the extraction of hadron matrix elements \cite{Hackett:2024nbe}.
In the following, we focus on the nucleon sector, where detailed numerical studies are complemented by effective field theory analyses. This combination provides a particularly clear setting to illustrate and support the general mechanism of current-enhanced ESC introduced in Ref.~\cite{Barca:2025det}.
\subsection{Nucleon matrix elements}
Specialising the general framework discussed above to the nucleon sector, the corresponding three-point function reads
\begin{align}
\label{nucleon3pt}
C_{\rm 3pt}(\vec{p}', t; &\vec{q}, \tau)
= ~
\langle \mathrm{O}_{N}(\vec{p}', t)~\mathcal{J}(\vec{q}, \tau)~\bar{\mathrm{O}}_{N}(\vec{p}, 0) \rangle
\\
=&~
\nonumber
Z^{|N(\vec{p}')\rangle}_{\mathrm{O}_{N}}
\bar{Z}^{|N(\vec{p})\rangle}_{\mathrm{O}_{N}}
~\langle N(\vec{p}')|\mathcal{J}(\vec{q})|N(\vec{p})\rangle
~ e^{-E'_{N}(t-\tau)} ~e^{-E_{N}\tau}
~+
\\
\label{nucleon3pt_2}
+&~
\sum_{(n,m) \neq (N,N)}
Z^{|n(\vec{p}')\rangle}_{\mathrm{O}_{N}}
\bar{Z}^{|m(\vec{p})\rangle}_{\mathrm{O}_{N}}
~\langle n(\vec{p}')|\mathcal{J}(\vec{q})|m(\vec{p})\rangle
~ e^{-E'_n(t-\tau)} ~e^{-E_m\tau}~
\quad (\text{ESC})
~,
\end{align}
where the (nucleon) ground-state contribution has been isolated from the ESC.
The first term contains the nucleon matrix element of interest, while the remaining terms encode all other transitions ($N\pi \to N\pi$, $N\to N\pi$, ...). The corresponding nucleon two-point function reads
\begin{equation}
\label{nucleon2pt}
C_{\rm 2pt}(\vec{p}, t)
=
\langle \mathrm{O}_{N}(\vec{p}, t)~\bar{\mathrm{O}}_{N}(\vec{p}, 0)
=
|Z^{|N(\vec{p})\rangle}_{\mathrm{O}_{N}}|^2~ e^{-E_N t}
~+~
\sum_{n\neq N} |Z^{|n(\vec{p})\rangle}_{\mathrm{O}_{N}}|^2 ~e^{-E_n t}.
\end{equation}
To extract the matrix element, we construct the standard ratio
\begin{equation}
\label{ratiomethod}
R^{\mathcal{J}}(\vec{p}', t; \vec{q}, \tau) = \frac{C^{\mathcal{J}}_{\rm 3pt}(\vec{p}', t; \vec{q}, \tau)}{C_{\rm 2pt}(\vec{p}', t)}
\sqrt{
\frac{C_{\rm 2pt}(\vec{p}', \tau) ~C_{\rm 2pt}(\vec{p}', t) ~C_{\rm 2pt}(\vec{p}, t-\tau)}
{C_{\rm 2pt}(\vec{p}, \tau) ~C_{\rm 2pt}(\vec{p}, t) ~C_{\rm 2pt}(\vec{p}', t-\tau)}}~.
\end{equation}
Inserting Eqs.~\eqref{nucleon3pt_2}–\eqref{nucleon2pt} shows that the term proportional to 
$\langle N(\vec{p}')|\mathcal{J}(\vec{q})|N(\vec{p})\rangle$ is independent of $t$ and $\tau$. 
Residual time dependence therefore provides a direct measure of ESC.

In the following, we illustrate this mechanism in several nucleon channels and kinematic setups where excited-state effects are particularly pronounced.

\subsubsection{Excited-state contamination in nucleon pseudoscalar and axial-vector matrix elements}
\begin{figure}[tbp]
  \centering
\begin{subfigure}{1.\linewidth}
    \centering
    \includegraphics[width=\linewidth]{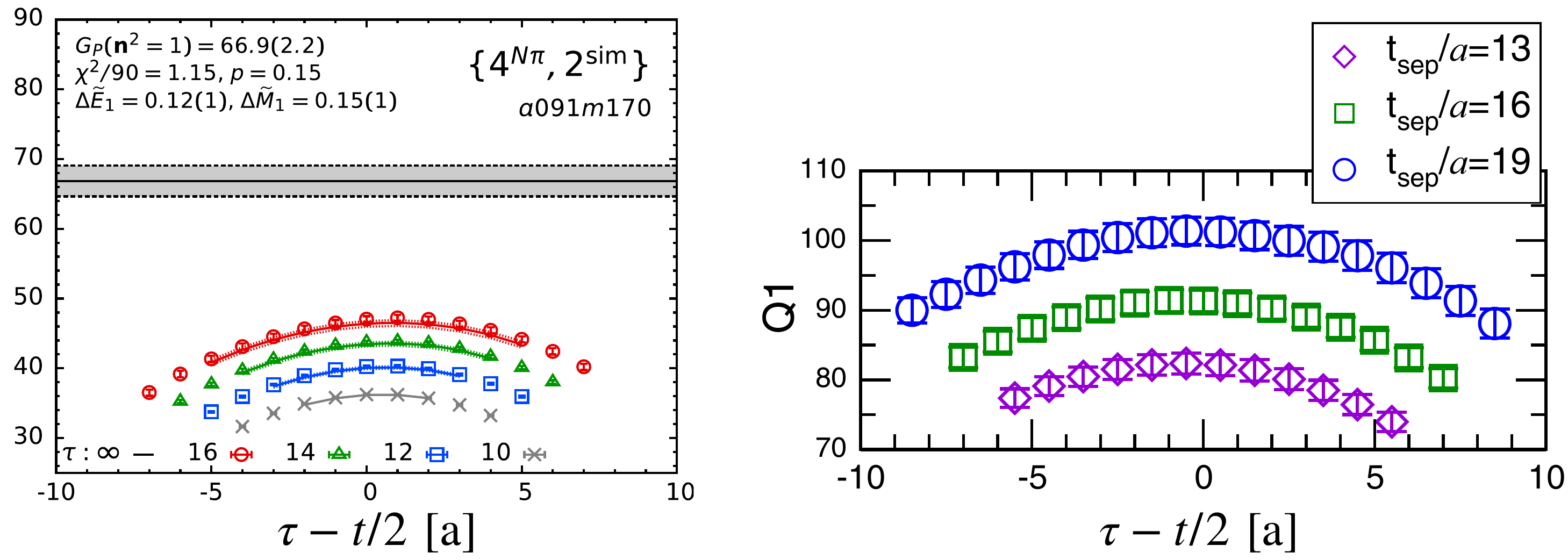}
    \label{fig:top-left}
  \end{subfigure}\hfill
  \vspace{-0.5em}
\begin{subfigure}{0.54\linewidth}
    \centering
    \includegraphics[width=\linewidth]{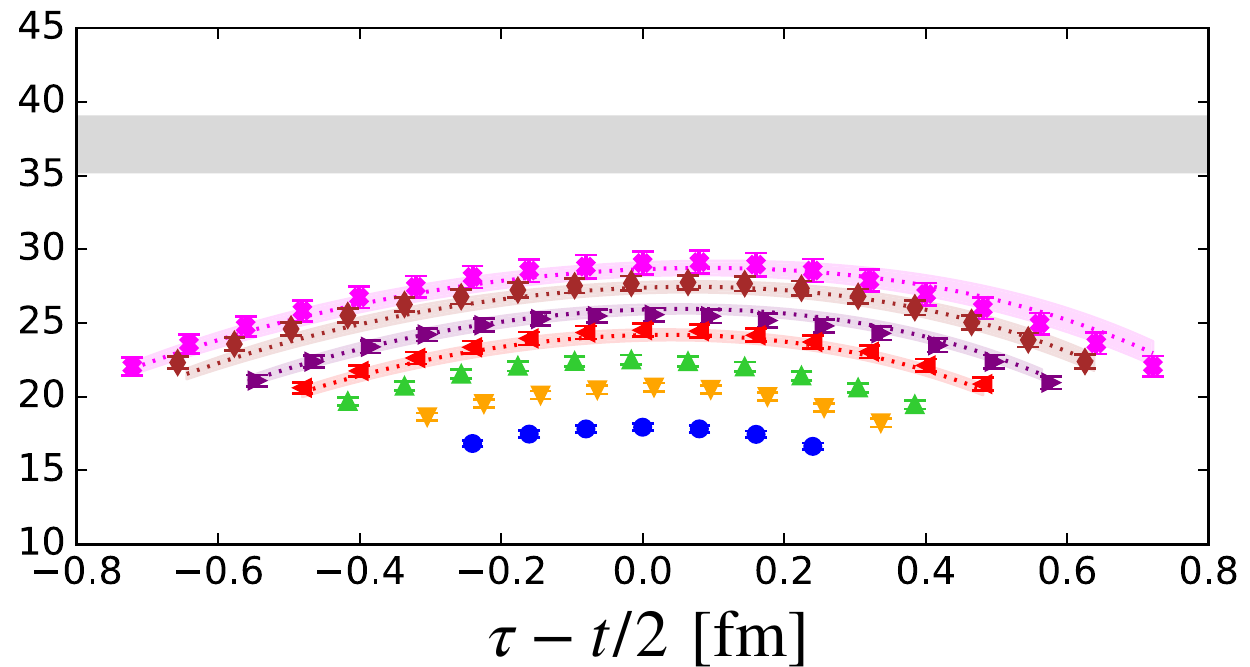}
    \label{fig:bottom}
  \end{subfigure}

  \caption{Standard ratios in Eq.~\eqref{ratiomethod} with a pseudoscalar current insertion at non-zero momentum transfer and different source-sink separations from different lattice groups.
  (top left): Results from the NME Collaboration, taken from Ref.~\cite{Park:2021ypf}. 
  The data are $t$- and $\tau$-dependent at different source-sink separations (different colours) 
  and they differ from the expected nucleon matrix element (grey band), extracted using multi-state fits.
  (top right) and (bottom): Results from the PACS Collaboration \cite{Aoki:2025taf} and ETM Collaboration \cite{Alexandrou:2020okk}, respectively, exhibit similar large ESC.
  }
  \label{fig:ratio_g4}
\end{figure}

Inserting the isovector pseudoscalar bilinear current $\mathcal{P}(\tau)=\bar{u}(\tau) \gamma_5 d(\tau)$ allows one to extract $\langle N(\vec p')|\mathcal P(\vec q)|N(\vec p)\rangle$, which is proportional to the nucleon pseudoscalar form factor $G_{\mathcal P}$.
The corresponding ratios at $\vec p'=\vec 0$, $\vec p=-\vec q$ (Fig.~\ref{fig:ratio_g4}) show a pronounced dependence on both $t$ and $\tau$. The absence of a plateau and the significant deviation from the ground-state expectation demonstrate that ESC is very large.

\begin{figure}[tbp]
  \centering

\begin{subfigure}{1.\linewidth}
    \centering
    \includegraphics[width=\linewidth]{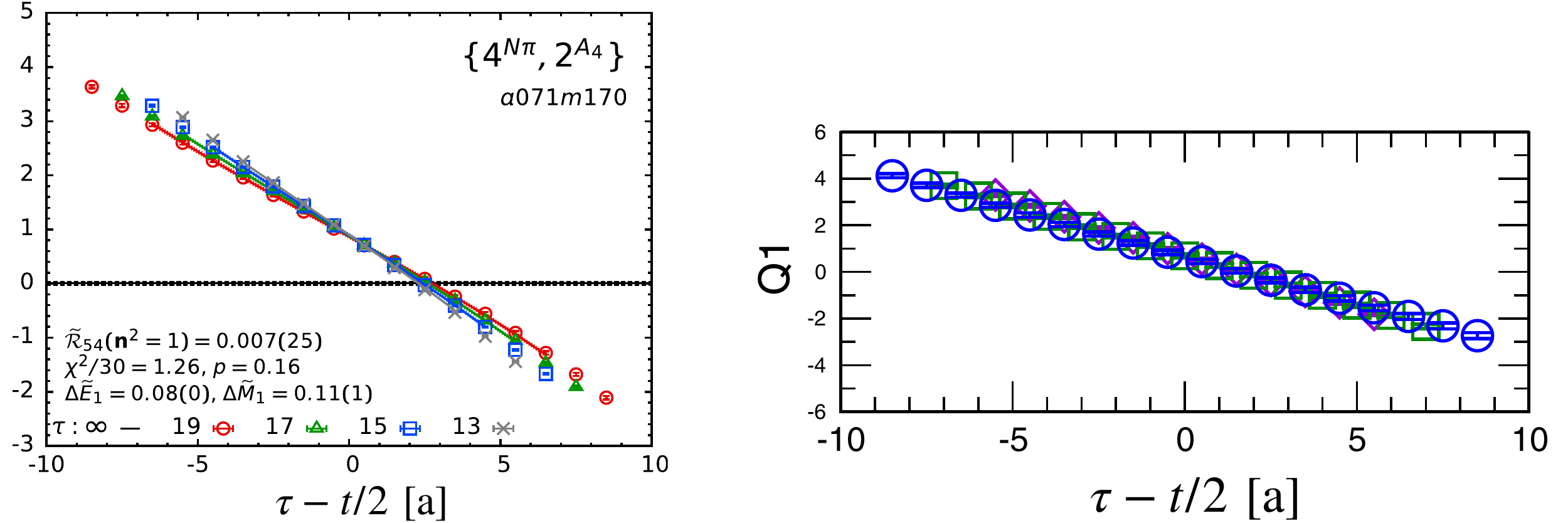}
  \end{subfigure}

  \vspace{0.7em}

\begin{minipage}{0.42\linewidth}
    \centering
    \includegraphics[width=\linewidth]{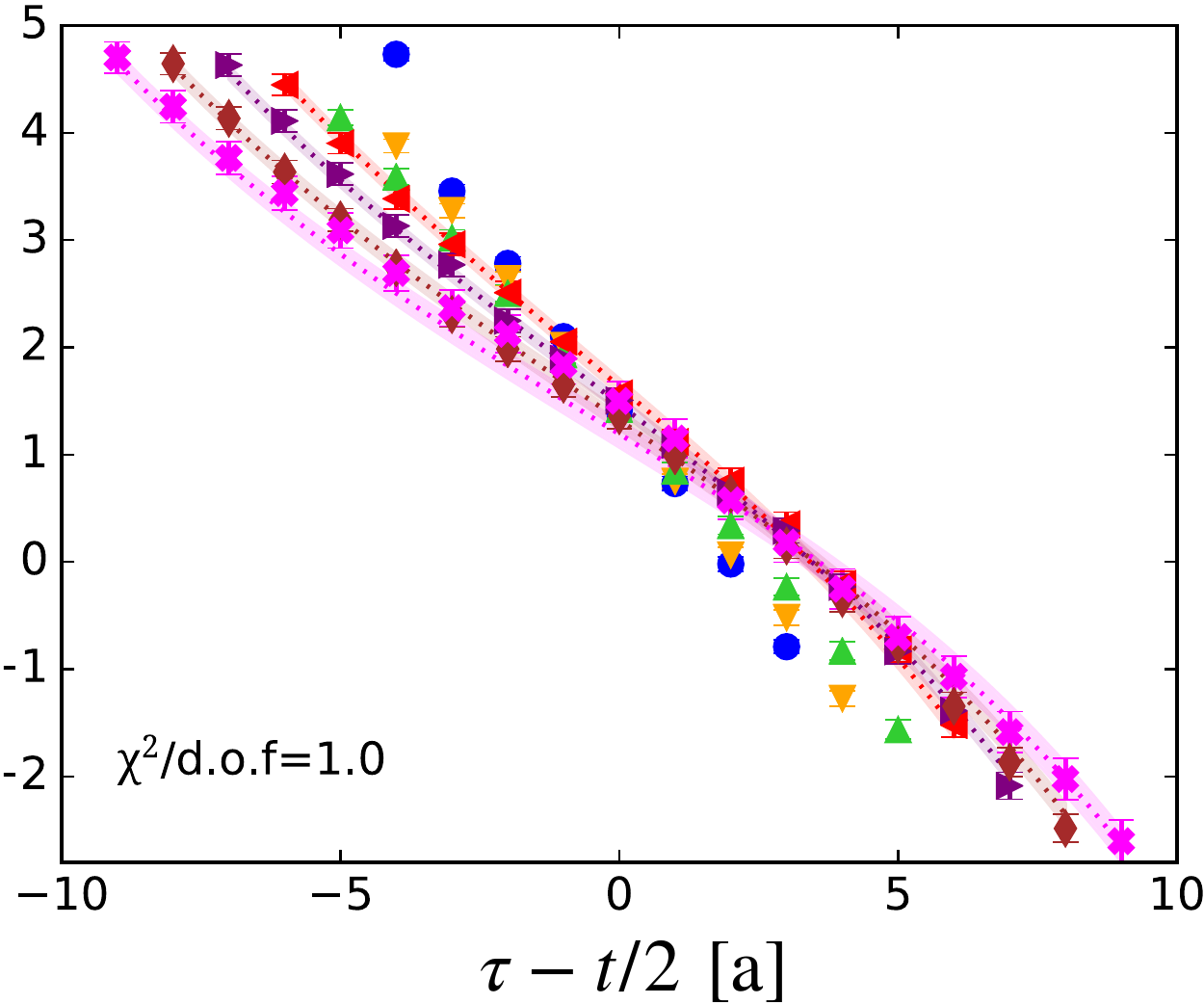}
  \end{minipage}\hfill
  \begin{minipage}{0.48\linewidth}
    \captionsetup{type=figure}
    \caption{
    Ratios in Eq.~\eqref{ratiomethod} with a temporal axial-vector 
    current insertion $\mathcal{J}=\bar{u} \gamma_4 \gamma_5 d$
    at non-zero momentum transfer and different
    source-sink separations from different lattice groups.
    (top left): Results from the NME Collaboration,
    Ref.~\cite{Park:2021ypf}.
    (top right) and (bottom): PACS
    \cite{Aoki:2025taf} and ETM
    \cite{Alexandrou:2020okk}, respectively,
    show similarly large ESC.
    The grey band in the top left plot highlights
    the extracted nucleon matrix element.
    }
    \label{fig:ratio_g4g5}
  \end{minipage}
\end{figure}

An even stronger effect is seen for the temporal axial current $\mathcal A_4(\tau)=\bar u(\tau)\gamma_4\gamma_5 d(\tau)$, whose matrix element determines the induced pseudoscalar $\widetilde G_P$ and axial-vector $G_A$ form factors.
The corresponding ratios (Fig.~\ref{fig:ratio_g4g5}) exhibit an almost linear dependence on $\tau$, indicating a dominant ESC. Simulation details can be found in Refs.~\cite{Park:2021ypf,Aoki:2025taf,Alexandrou:2020okk}.

ChPT provides crucial insight.
At leading order (LO-ChPT), $N\pi$ states contribute only at the few-percent level to nucleon two-point functions at near-physical pion masses, while $N\pi\pi$ states are further suppressed~\cite{Bar:2015zwa,Bar:2018wco,Tiburzi:2015tta}. 
For three-point functions, however, LO-ChPT predicts qualitatively different behaviour: 
$N\pi$ contributions can reach ${\mathcal O}(40\%)$ at $t\sim2\,\mathrm{fm}$ and depend on $Q^2$~\cite{Bar:2018xyi,Bar:2019gfx}. 
The dominant terms arise when the pion carries the momentum injected by the current; these contributions are not volume suppressed and can therefore remain sizable at accessible source–sink separations.

Using these ChPT insights, and in particular that the $N\pi$ contributions dominate both the pseudoscalar and axial-vector channels, in Ref.~\cite{Park:2021ypf} they constraint the excited-state energy in the $\mathcal{P}$ and $\mathcal{A}_k$ channels by extracting the excite-state ($N\pi$) energy from the $\mathcal{A}_4$ channel.

In Ref.~\cite{RQCD:2019jai}, the ChPT predictions of Refs.~\cite{Bar:2018xyi, Bar:2019gfx} have been generalised to different moving frames, 
and the expressions for the $N\pi$ contribution have been used to constraint the first excited state at source ($\propto \langle N | \mathcal{J}|N\pi\rangle$) and sink ($\propto \langle N\pi |\mathcal{J}|N\rangle$) via a ChPT-inspired ansatz. 

An important consistency check is provided by the PCAC relation $\partial^\mu \mathcal{A}_\mu = 2i m_\ell \mathcal{P}$,
with $m_\ell$ the light quark mass, which implies
\begin{equation}
m_N G_A(Q^2) = m_\ell G_P(Q^2) + \frac{Q^2}{2m_N}\tilde{G}_P(Q^2).
\end{equation}
On the lattice, this relation, which is the generalisation of the Goldberger-Treimann relation \cite{PhysRev.111.354}, can be broken by discretisasion effects, or from an improper treatement of ESC in the extraction of the nucleon form factors from nucleon three-point functions.
Nucleon form factors extracted from standard multi-state fits were found to violate this relation well beyond expected discretisation effects~\cite{Gupta:2017dwj,Ishikawa:2018rew,Liang:2018pis,Bali:2018qus}. 

In Ref.~\cite{RQCD:2019jai}, it was demonstrated that if the $N\pi$ contributions are not properly accounted for, violations of the Goldberger–Treiman relation--also known in the literature as PCAC relation at the level of form factors--of up to $\sim 40\%$ are observed at small momenta and near-physical pion masses. 
In contrast, incorporating the ChPT-inspired $N\pi$ contribution restores the relation at the few-percent level~\cite{RQCD:2019jai}, consistent with residual lattice artifacts. Independent determinations confirm this behaviour~\cite{Jang:2019vkm,Park:2021ypf}.
These results demonstrate that a precise systematic treatment of ESC is therefore essential and must be extended beyond the nucleon case to any hadron three-point functions.
This is particularly important in precision studies of rare decays~\cite{Erben:2025jkq}, where lattice inputs for hadronic matrix elements enter directly in searches for physics beyond the Standard Model (BSM).

\subsubsection{Excited-state contamination in nucleon scalar matrix elements}
\begin{figure}[tbp]
  \centering

\begin{subfigure}{1.\linewidth}
    \centering
    \includegraphics[width=1\linewidth]{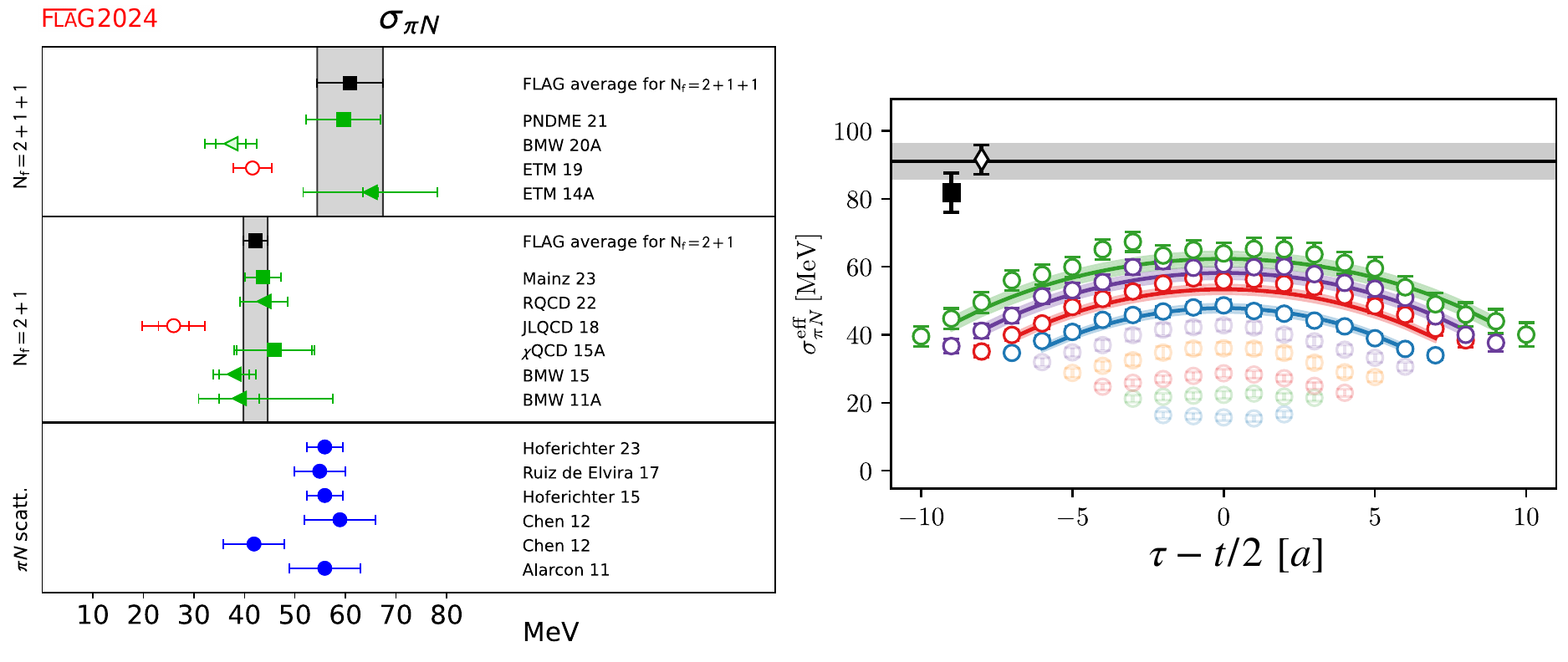}
  \end{subfigure}\hfill  
  \vspace{0.5em}

\begin{subfigure}{0.95\linewidth}
    \centering
    \includegraphics[width=\linewidth]{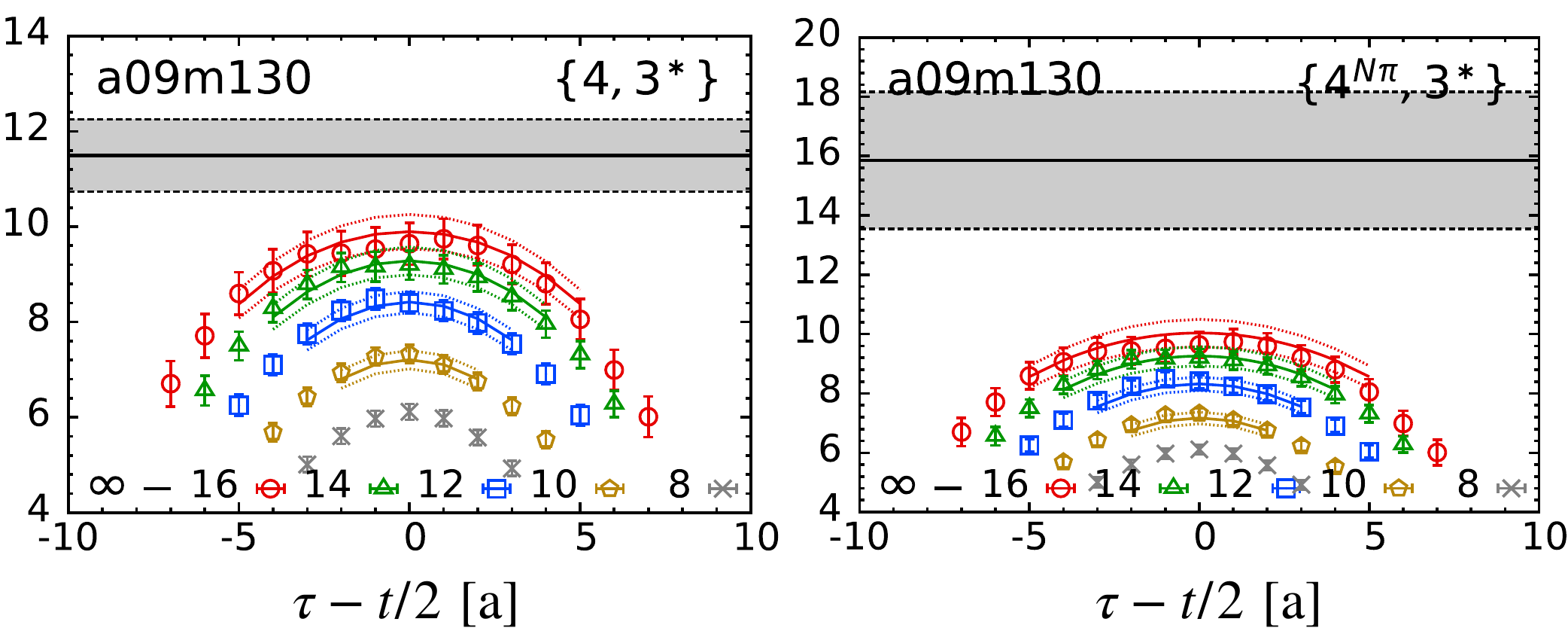}
  \end{subfigure}\hfill
    \caption{
    (top left) Comparison of lattice determinations of $\sigma_{\pi N}$
    with $N_f=2+1$ and $N_f=2+1+1$ and phenomenological determinations from 
    nucleon-pion scattering data. The grey band and black data represent
    the FLAG 2024 average.
    (top right) Lattice determination from the Mainz lattice group of $\sigma_{\pi N}$ 
    on a single ensemble at $m_\pi = 200~\rm MeV$ at different source-sink separations $t$
    (different colours). The grey band highlights the fitted value for $\sigma_{\pi N}$ and the 
    black diamond is the result of a two-state fit to the summed correlator. For more details see \cite{PhysRevLett.131.261902}.
    (bottom) Ratios in Eq.~\eqref{ratiomethod} with an isoscalar scalar current 
    insertion at zero momentum transfer and different source-sink separations \cite{Gupta:2021ahb}.
    }
    \label{fig:ratio_I}
\end{figure}

Nucleon scalar matrix elements are of particular phenomenological interest. 
The isoscalar matrix element $\langle N | \bar q q | N\rangle$ defines the quark sigma terms, 
$\sigma_{qN}$, and in particular $\sigma_{\pi N}=\sigma_{uN}+\sigma_{dN}$.
These quantities enter the nucleon mass decomposition and provide essential input for BSM searches, 
such as dark-matter-nucleon interactions in Higgs portal scenarios.
Phenomenological analyses based on pion-nucleon scattering report
$\sigma_{\pi N}\approx 59~\mathrm{MeV}$.
First-principles determinations from lattice QCD are systematically improvable,
yet most lattice results tend to yield smaller values, leading to a tension between 
phenomenology and lattice determinations, see Fig.~\ref{fig:ratio_I}.

In this channel, ESC can be particularly severe.
As shown in the top right and bottom plots in Fig.~\ref{fig:ratio_I}, the ratios exhibit a pronounced dependence on $t$ and $\tau$.
A particularly instructive example is provided in Ref.~\cite{Gupta:2021ahb}, where two well-motivated fit strategies were compared 
to the same data, see bottom plot in Fig.~\ref{fig:ratio_I}.
In the so-called $\{4,3^\ast\}$ fit, wide priors are imposed on the excited-state energies to stabilise the analysis, while the $\{4_{N\pi},3^\ast\}$ fit constrains the first excited state to lie close to the non-interacting energy of the lowest positive-parity multi-hadron states, $N(\vec{1})\pi(-\vec{1})$ or $N(\vec{0})\pi(\vec{0})\pi(\vec{0})$. The latter choice is motivated by their ChPT analysis, which predicts sizable contributions from $N\pi$ and $N\pi\pi$ states in this channel.

Although both fits describe the data equally well and yield comparable $\chi^2$, the extracted values of the isoscalar scalar charge differ significantly.
For the near-physical ensemble ($M_\pi = 138\,\mathrm{MeV}$), the two strategies lead to $\sigma_{\pi N} \approx 40\,\mathrm{MeV}$ with the $\{4,3^\ast\}$ fit and $\approx 60\,\mathrm{MeV}$ with the $\{4_{N\pi},3^\ast\}$ fit, respectively. 
The latter is consistent with the phenomenological values, see "PNDME 21" result in the top left plot of Fig.~\ref{fig:ratio_I}. 
In Ref.~\cite{Gupta:2021ahb}, they find that at heavier pion masses ($M_\pi \approx 315\,\mathrm{MeV}$), the effect of the $N\pi$ and $N\pi\pi$ states is difficult to resolve, while at the physical point it is clearly visible. 

An explanation for this $m_\pi$ dependence is related to the structure of the lowest scalar excitations.
At sufficiently heavy pion masses, the $\sigma$ meson becomes stable, and the isoscalar scalar current couples predominantly to this single-particle state. Our variational analysis at $M_\pi = 429\,\mathrm{MeV}$, discussed in the next section, supports this interpretation.
At lighter pion masses, however, the $\sigma$ becomes unstable and the scalar current strongly couples to $\pi\pi$ S-wave states, consistent with the ChPT prediction in Ref.~\cite{Gupta:2021ahb}. Our preliminary analysis at $m_\pi = 222~\rm MeV$ supports also this interpretation.
The identification of the leading contributing states is therefore not universal, but depends on the spectrum, kinematics and choice of operators and currents.

\subsection{Variational analyses for nucleon matrix elements}
\subsubsection{Nucleon pseudoscalar and axial-vector channels}
Motivated by the LO-ChPT predictions of Refs.~\cite{Bar:2018xyi,Bar:2019gfx,RQCD:2019jai}, a variational analysis including $N$- and the lowest $N\pi$-like interpolating operators was performed in Refs.~\cite{Barca:2022uhi,Barca:2021iak}. 
This pilot study was carried out on an ensemble with $m_\pi = 429~\mathrm{MeV}$ and $N_f=3$.

In this approach, the $N\to N\pi$ and $N\pi \to N$ transition contributions were computed explicitly and incorporated into a generalized eigenvalue problem (GEVP) analysis, discussed in more details in the next section. The resulting GEVP-improved nucleon operators effectively suppress the leading $N\pi$ contamination in the three-point functions.

\begin{figure}[tbp]
  \centering
    \includegraphics[width=1.0\linewidth]{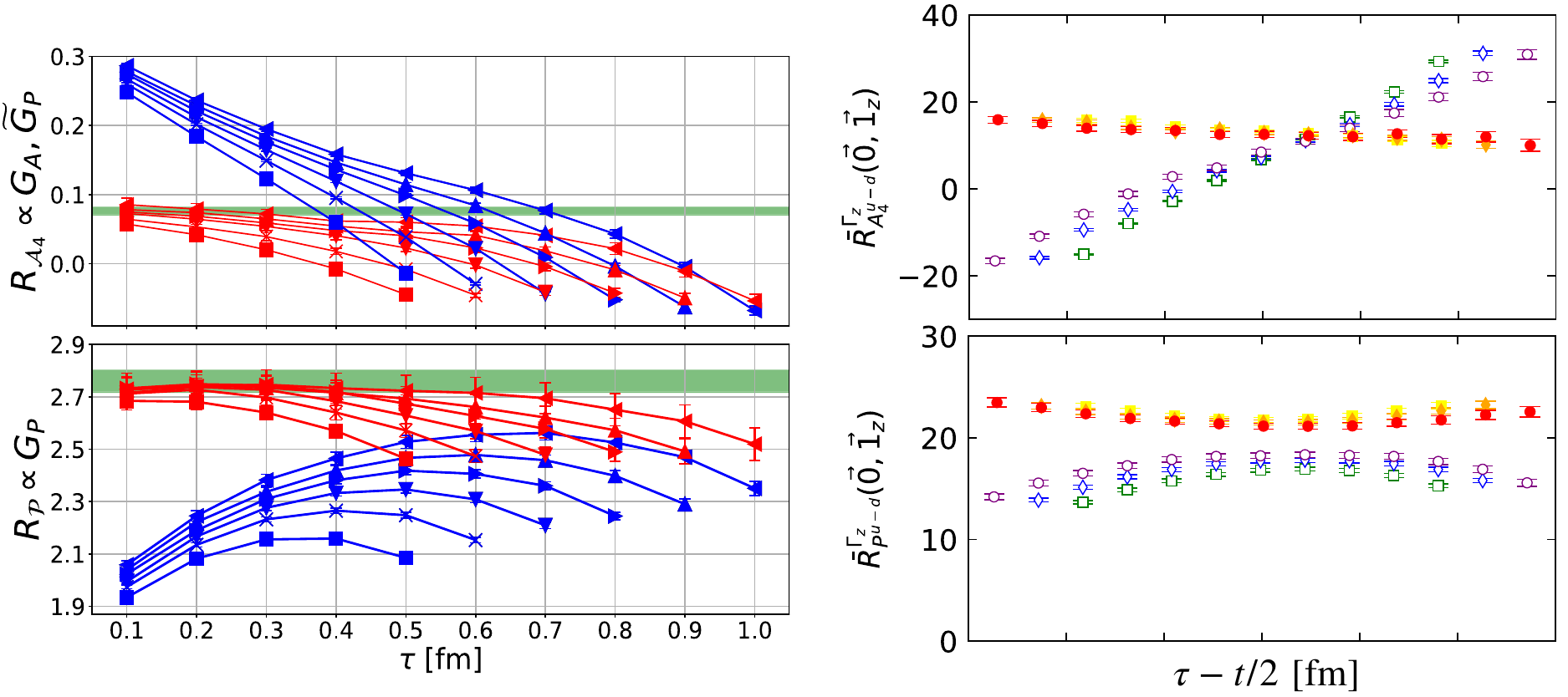}
    \caption{
    GEVP-improved ratios with a pseudoscalar and temporal
    axial-vector current insertion at non-zero momentum transfer 
    and different source-sink separations. 
    (left) Results from Ref.~\cite{Barca:2022uhi} at $m_\pi = 429~\rm MeV$.
    (right) Results from the ETM collaboration in Ref.~\cite{Alexandrou:2024tin} at $m_\pi = 346~\rm MeV$.
    }
    \label{fig:gevp_ratios_g4g5_g5}
\end{figure}

As shown in the left panel of Fig.~\ref{fig:gevp_ratios_g4g5_g5}, ratios constructed with GEVP-improved nucleon operators (red symbols) approach the ChPT-inspired ground-state expectation (green band), in contrast to those obtained with standard nucleon operators (blue symbols). 
This improvement is observed across source–sink separations $6 \leq t/a \leq 11$ ($a=0.098~\mathrm{fm}$), demonstrating that the $N\pi$ state provides the dominant ESC in these channels, consistent with LO-ChPT predictions even at $m_\pi = 429~\mathrm{MeV}$.

These findings were subsequently confirmed by the ETM collaboration at $m_\pi = 346~\mathrm{MeV}$ and at the physical-pion-mass ensemble ($m_\pi = 131~\mathrm{MeV}$)~\cite{Alexandrou:2024tin}. 
In the right panel of Fig.~\ref{fig:gevp_ratios_g4g5_g5}, a clear reduction of excited-state effects is observed when employing GEVP-improved operators (filled symbols) in the pseudoscalar and temporal axial channels.
In contrast, applying the same variational strategy to other nucleon matrix elements, such as vector, scalar, and tensor currents, does not lead to a comparable improvement~\cite{Alexandrou:2024tin}. 
The persistence of sizable ESC in these channels indicates that the dominant contributing states are not the lowest $N\pi$ states.

This can be understood within the framework of current-enhanced states introduced in Ref.~\cite{Barca:2025det}. 
In these channels, different multi-hadron states are preferentially enhanced by the structure of the inserted current and the chosen kinematics. 
The leading ESC therefore depends not only on the energy spectrum but also on the quantum numbers of the current and kinematics.
\subsubsection{Nucleon scalar channel}
Motivated by the current-enhancement mechanism and the findings of Ref.~\cite{Alexandrou:2024tin}, we performed a variational analysis of the isoscalar scalar matrix element using $N$- and $N\sigma$-like interpolating operators ($\sigma \sim \bar{u}u + \bar{d}d + \bar{s}s$) on an ensemble with $m_\pi=429~\rm MeV$, $N_f=3$, and $V/a^4=48\times 24^3$~\cite{Barca:2024hrl}. 

On this ensemble, the $\sigma$ meson is a stable, deeply bound state with mass $\approx 550~\rm MeV$. Solving the GEVP for the matrix of two-point functions reveals the nucleon ground state and a second state close to $m_N+m_\sigma$, which we interpret as the $N\sigma$ state. This state lies below the lowest $N(\vec{1})\pi(-\vec{1})$ and $N\pi\pi$ states.
\begin{figure}[t]
\centering
\begin{minipage}[t]{0.62\linewidth}\vspace{0pt}
  \centering
  \includegraphics[width=\linewidth]{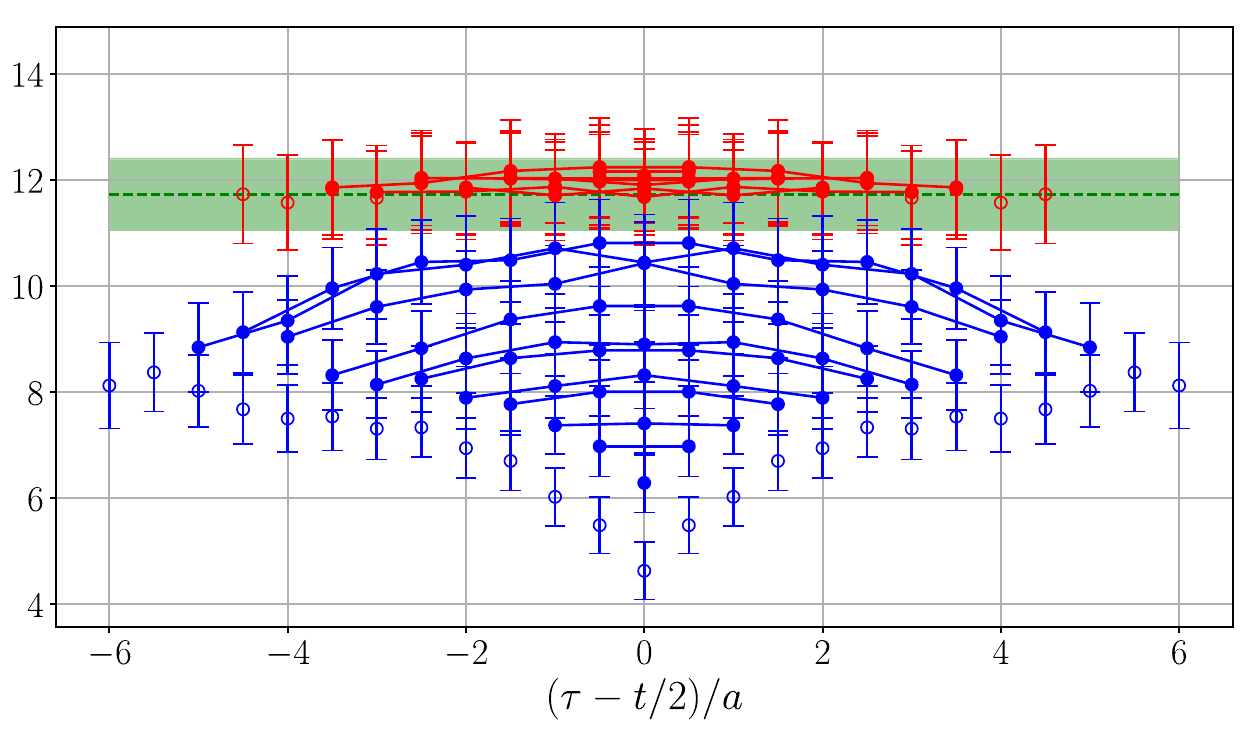}
\end{minipage}\hspace{0.03\linewidth}\begin{minipage}[t]{0.35\linewidth}\vspace{3pt}
\captionof{figure}{Comparison between standard (blue) and GEVP-improved (red) ratios in Eq.~\eqref{ratiomethod}
with the isoscalar scalar current at zero momentum and source-sink separations, 
and at $m_\pi = 429~\rm MeV$, see Ref.~\cite{Barca:2024hrl}.
The ESC is completely removed within statistical uncertainties.
The dominant source of ESC is the $N\sigma$ state. 
}
\label{fig:gevp_ratio_I}
\end{minipage}
\end{figure}
Using the nucleon components of the GEVP eigenvectors, we construct an improved interpolating operator that removes the contributions proportional to $\langle N\sigma | \mathcal{S}^{u+d, s} | N \rangle$ and $\langle N | \mathcal{S}^{u+d, s} | N\sigma \rangle$ from the scalar three-point function.
The resulting GEVP ratios, shown in Fig.~\ref{fig:gevp_ratio_I}, become constant within uncertainties at all source–sink separations $t>0.4~\rm fm$, indicating that there was only one dominant ESC ($N\sigma$) and that it has been completely removed.
This improvement reflects that the $N\sigma$ state is both the lowest scalar excitation in the rest frame and \textit{current-enhanced}.

At lighter pion masses the situation changes qualitatively.
In a preliminary analysis at $m_\pi=222~\rm MeV$ ($N_f=2+1$, $V/a^4=96\times48^3$), the $\sigma$ becomes unstable and the scalar channel is dominated by $\pi\pi$ S-wave states with energy $\approx 420~\rm MeV$, below the $\sigma$ resonance.
The second GEVP state is consistent with a non-interacting $N\pi\pi$ level, though mixing with $N\pi$ states cannot be excluded because of the quasi-degeneracy.
Despite this more intricate structure, constructing GEVP-improved operators significantly reduces the ESC in the three-point functions. These encouraging results suggest that a variational basis including $N$- and $N\sigma$-like operators remains beneficial also at lighter pion masses. A dedicated analysis at near-physical pion masses will be necessary to fully assess the interplay between $N\sigma$, $N\pi$, and $N\pi\pi$ contributions.

\subsubsection{Nucleon vector channel}
In Ref.~\cite{Barca:2025det}, nucleon isovector vector matrix elements were studied using a variational basis with $N$- and $N\rho$-like interpolating operators on an ensemble with $m_\pi=429~\rm MeV$. 
At first sight this choice appears counter-intuitive, since $N\rho$ states are substantially heavier than the $N\sigma$, $N\pi$, or $N\pi\pi$ states, particularly in the rest frame. 
While heavier states are exponentially suppressed at sufficiently large Euclidean times, current lattice simulations are restricted to moderate source-sink separations, and if these heavy states appear with a large prefactor in the spectral decomposition, they can contribute significantly to the three-point function despite its larger energy. 

The variational analysis was performed on the same ensemble as in Refs.~\cite{Barca:2022uhi, Barca:2024hrl}, with $m_\pi=429~\rm MeV$, $N_f=3$, and $V/a^4=48\times 24^3$. 
The $\rho$ two-point function exhibits an effective mass of $\approx 860~\rm MeV$ in the rest frame, below the lowest $\pi(\vec{1})\pi(-\vec{1})$ level. 
Solving the GEVP in both rest and moving frames yields a second state close to the non-interacting $N\rho$ level. 
Constructing GEVP-improved nucleon operators using the nucleon components of the eigenvectors leads to a substantial reduction of ESC, particularly in moving frames (see Appendix of Ref.~\cite{Barca:2025det}).

Taken together, these results establish a coherent pattern:  
in the isovector pseudoscalar and axial-vector channels the dominant ESC originates from $N\pi$ states;  
in the isoscalar scalar channel from $N\sigma$ states; and in the isovector vector channel from $N\rho$ states.  
The dominant contributions are therefore determined primarily by the structure of the inserted current and the kinematics, rather than solely by the ordering of the energy spectrum. 
These findings provide direct numerical support for the current-enhancement mechanism proposed in Ref.~\cite{Barca:2025det} and discussed in the next section.

 \section{Mechanism of current-enhancement}
We now discuss how specific excited-state contributions can become enhanced due to the choice of current and kinematics, following Ref.~\cite{Barca:2025det}. Consider a basis of interpolating operators with the quantum numbers of a hadron $H$,
\begin{equation}
    \label{basisH}
    \mathbb{B}_H = \{ \mathrm{O}_H, \mathrm{O}_{HM}, \mathrm{O}_{HMM}, ...\},
\end{equation}
where $\mathrm{O}_{HM}$ and $\mathrm{O}_{HMM}$ denote operators which overlap to multi-particle states.
There is a tower of interpolating operators to be considered in the basis to account for the infinite tower of states in the spectrum of $H$.
From this basis one constructs the matrix of two-point functions
\begin{equation}
    C_{ij}(\vec{p}, t) = \langle \mathrm{O}_i(\vec{p}, t)~\bar{\mathrm{O}}_j(\vec{p}, 0) \rangle
    \qquad \qquad \text{with} ~\mathrm{O}_i, \mathrm{O}_j \in \mathbb{B}_H,
\end{equation}
and determine the matrix of eigenvalues $\Lambda=\mathrm{diag}\left(\lambda^H, \lambda^{HM}, \lambda^{HMM}, ...\right)$ and eigenvectors $V=\left(\vec{v}^H, \vec{v}^{HM}, \vec{v}^{HMM}, ...\right)$ by solving the GEVP
\begin{equation}
    C(\vec{p}, t) V(\vec{p}, t; t_0) = C(\vec{p}, t_0) \Lambda(\vec{p}, t; t_0) V(\vec{p}, t; t_0).
\end{equation}
For a complete basis, the eigenvectors isolate the energy eigenstates and become time-independent. 
The ground-state (H) components of $V$ can be used to construct a GEVP-improved interpolator,
\begin{equation}
    \mathrm{\Omega}_H = \sum_{\mathrm{O}_i \in \mathbb{B}_H} v_{\mathrm{O}_i}^H \mathrm{O}_i~.
\end{equation}
These improved operators can be adopted to construct GEVP-improved two-point functions
\begin{equation}
\label{c2pt_gevp}
\widetilde{C}_{2pt}(\vec{p}, \vec{t})
=
~\langle \mathrm{\Omega}_H(\vec{p}, t) ~ \bar{\mathrm{\Omega}}_H(\vec{p}, 0)\rangle
=
~\sum_{\mathrm{O}_i \in \mathbb{B}_H}
|v^H_{\mathrm{O}_i}(\vec{p})|^2~
\langle
\mathrm{O}_i(\vec{p}, t)~ \bar{\mathrm{O}}_j(\vec{p}, 0)\rangle,
\end{equation}
and GEVP-improved three-point functions
\begin{align}
\label{c3pt_gevp}
\widetilde{C}_{3pt}(\vec{p}', \vec{t}; \vec{q}, \tau)
=&
~\langle \mathrm{\Omega}_H(\vec{p}', t)~\mathcal{J}(\vec{q}, \tau)~ \bar{\mathrm{\Omega}}_H(\vec{p}, 0)\rangle
\\
\label{c3pt_gevp2}
= &
~\sum_{\mathrm{O}_i, \mathrm{O}_j \in \mathbb{B}_H}
v^H_{\mathrm{O}_i}(\vec{p}')~
v^H_{\mathrm{O}_j}(\vec{p})~
\langle \mathrm{O}_i(\vec{p}', t)~\mathcal{J}(\vec{q}, \tau)~ \bar{\mathrm{O}}_j(\vec{p}, 0)\rangle.
\end{align}
For a complete basis, the ESC in the GEVP-improved correlation functions is completely removed, and one can see the ground state dominance of the $H$ hadron. These GEVP-improved correlation functions can then be used to construct ratios equivalent to Eq.~\eqref{ratiomethod}. 

The key observation concerns the volume scaling of the terms in Eq.~\eqref{c3pt_gevp2}: 
$v^{H}_{\mathrm{O}_{HM}}$ is expected to be suppressed by the spatial volume compared to $v^{H}_{\mathrm{O}_{H}}$.
However, the associated three-point function
\begin{equation}
    \langle \mathrm{O}_{HM}(\vec p',t)\,\mathcal J(\vec q,\tau)\,\bar{\mathrm{O}}_H(\vec p,0)\rangle,
\end{equation}
can contain Wick contractions that are enhanced by a (spatial) volume factor.
In particular, quark-line disconnected topologies allow the current-meson insertion and nucleon propagation to occur independently over the spatial volume. 
This volume enhancement can compensate the volume suppression of the eigenvector component, leading to an overall unsuppressed contribution.

For example, in the computation of $\langle \mathrm{O}_{N\pi}(\vec{p}', t)~\mathcal{J}(\vec{q}, \tau)~ \bar{\mathrm{O}}_N(\vec{p}, 0)\rangle$ for nucleon pseudoscalar and axial-vector matrix elements, there are quark-line connected $W_C$ and disconnected $W_D$ topologies that contribute to the Wick contractions and that read
\begin{align}
\label{W_C}
W_C
=&~
\langle
\sum_{\vec{x}_N, \vec{x}_\pi, \vec{z}}
e^{-i \vec{p}'_N \cdot \vec{x}_N}
e^{-i \vec{p}'_\pi \cdot \vec{x}_\pi}
e^{i \vec{q} \cdot \vec{z}}
~\langle \mathrm{O}_N(\vec{x}_N, t) \mathrm{O}_\pi(\vec{x}_\pi, t) \mathcal{J}(\vec{z}, \tau) \bar{\mathrm{O}}_N(\vec{p}, 0) \rangle_F~\rangle_G,
\\
W_D 
\label{W_D}
=&~
\langle
\sum_{\vec{x}_N, \vec{x}_\pi, \vec{z}}
e^{-i \vec{p}'_N \cdot \vec{x}_N}
e^{-i \vec{p}'_\pi \cdot \vec{x}_\pi}
e^{i \vec{q} \cdot \vec{z}}
~\langle \mathrm{O}_N(\vec{x}_N, t) \bar{\mathrm{O}}_N(\vec{p}, 0) \rangle_F ~\langle \mathrm{O}_\pi(\vec{x}_\pi, t) \mathcal{J}(\vec{z}, \tau) \rangle_F~\rangle_G.
\end{align}
In these expressions, $\vec{p}'_N + \vec{p}'_\pi=\vec{p}'$, $\langle \cdot \rangle_F$ represents the Wick contractions and $\langle \cdot \rangle_G$ the gauge average.
In the quark-line disconnected topology, the gauge average can be decomposed into a factorised (non-interacting) contribution and a gauge-connected contribution, where the latter encodes gluonic interactions between the nucleon-nucleon and pion-current blocks.
The non-interacting piece of the quark-line disconnected diagram reads
\begin{align}
W^{(0)}_D
=&~
C_{2pt}^N(\vec{p}'_N, t; \vec{p}, 0)
\times
\sum_{\vec{x}_\pi, \vec{z}}
e^{-i \vec{p}'_\pi \cdot \vec{x}_\pi}
e^{i \vec{q} \cdot \vec{z}}
~\langle \langle \mathrm{O}_\pi(\vec{x}_\pi, t) \mathcal{J}(\vec{z}, \tau) \rangle_F~\rangle_G,
\end{align}
where $C_{2pt}^N(\vec{p}'_N, t; \vec{p}, 0)$ denotes the nucleon two-point functions.
In the kinematics where $\vec{q}=\vec{p}'_\pi$, $\vec{p}'_N=\vec{p}$, and by using translational invariance, one gets
\begin{align}
W^{(0)}_D
=&~
C_{2pt}^N(\vec{p}, t)
\times
\sum_{\vec{x}_\pi, \vec{z}}
e^{i \vec{q} \cdot (\vec{z}-\vec{x}_\pi)}
~\langle \langle \mathrm{O}_\pi(\vec{x}_\pi, t) \mathcal{J}(\vec{z}, \tau) \rangle_F~\rangle_G
\\
(\vec{z}=\vec{x}_\pi+ \vec{r})
\quad
=&~
C_{2pt}^N(\vec{p}, t)
\times
(L/a)^3
\sum_{r}
e^{i \vec{q} \cdot \vec{r}}
~\langle \langle \mathrm{O}_\pi(\vec{0}, t) \mathcal{J}(\vec{r}, \tau) \rangle_F~\rangle_G~.
\end{align}
This shows that the quark-line disconnected diagrams are enhanced by a spatial volume factor, contrary to the quark-line connected diagrams.
In Ref.~\cite{Barca:2022uhi}, on the ensemble with $m_\pi=429~\mathrm{MeV}$ and $V/a^4=48\times 24^3$, we find that the quark-line disconnected diagrams are found to be two orders of magnitude larger than the quark-line connected diagrams at $\tau=t$.

This spatial volume enhancement of these three-point function comes when the pion carries the same momentum of the current, which is consistent with the LO-ChPT predictions \cite{Bar:2018xyi, Bar:2019gfx} and the many lattice QCD analyses discussed before.
Notice that $W_D^{(0)}$ contributes only when the nucleon-nucleon and the current-pion correlation functions are both non-zero.
For the spatial components of the axial-vector current, this is non-vanishing if the current has non-vanishing momentum along the direction of the axial-vector component, as $\langle \mathrm{O}_\pi(\vec{q})~ \mathcal{A}_k(\vec{q})\rangle \propto i q_k f_\pi$.
\begin{figure}[t]
\centering
  \includegraphics[width=1.\linewidth]{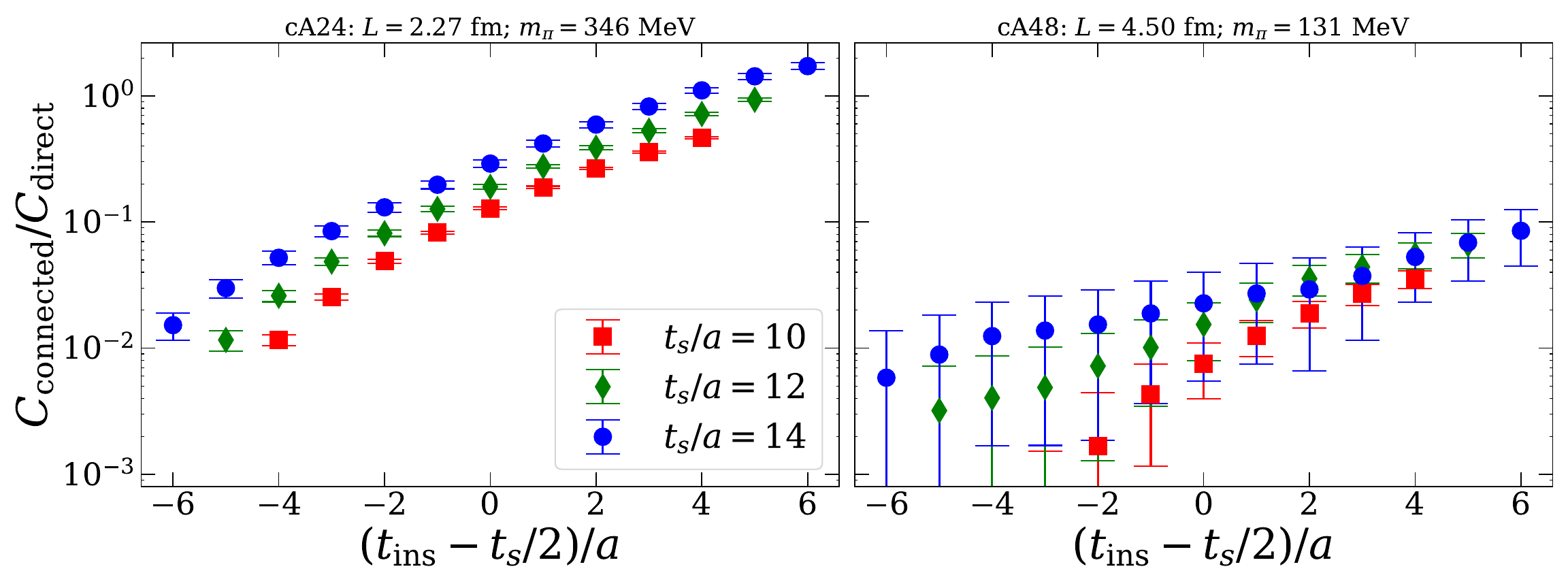}
\caption{
Ratio $W_C/W_D$ with the $N\pi$ interpolator at the source, cf. Eqs.~\eqref{W_C}-\eqref{W_D}.
The ratios are constructed with $\mathcal{J}=\mathcal{P}$ and at different source-sink separations $t_S$. 
(left) Ratios on ensemble "cA24" with $L=2.27~\rm fm$ and $m_\pi=346~\rm MeV$;
(right) Ratios computed on ensemble "cA48" with $L=4.50~\rm fm$ and $m_\pi = 131~\rm MeV$.
For more details, see Ref.~\cite{Alexandrou:2024tin}.
}
\label{fig:comparison_etmc_volume_scaling_ratios}
\end{figure}
In Ref.~\cite{Alexandrou:2024tin}, they study the ratio $W_C/W_D$ on two different ensembles, confirming that the ratio decreases approximately with $(L/a)^3$, and it's shown in Fig.~\ref{fig:comparison_etmc_volume_scaling_ratios}. A more clear comparison can be made at fixed pion masses and two different volumes as the amplitudes depend also on the pion masses, see Eq.~(2.45) of Ref.~\cite{RQCD:2019jai}.
The mechanism discussed above suggests concrete strategies to control large excited-state contamination in hadron three-point functions.
 \section{Strategies to account for large excited-state contamination}
\begin{figure}[tbp]
\centering
\begin{minipage}[tbp]{0.50\linewidth}\vspace{0pt}
  \centering
  \includegraphics[width=\linewidth]{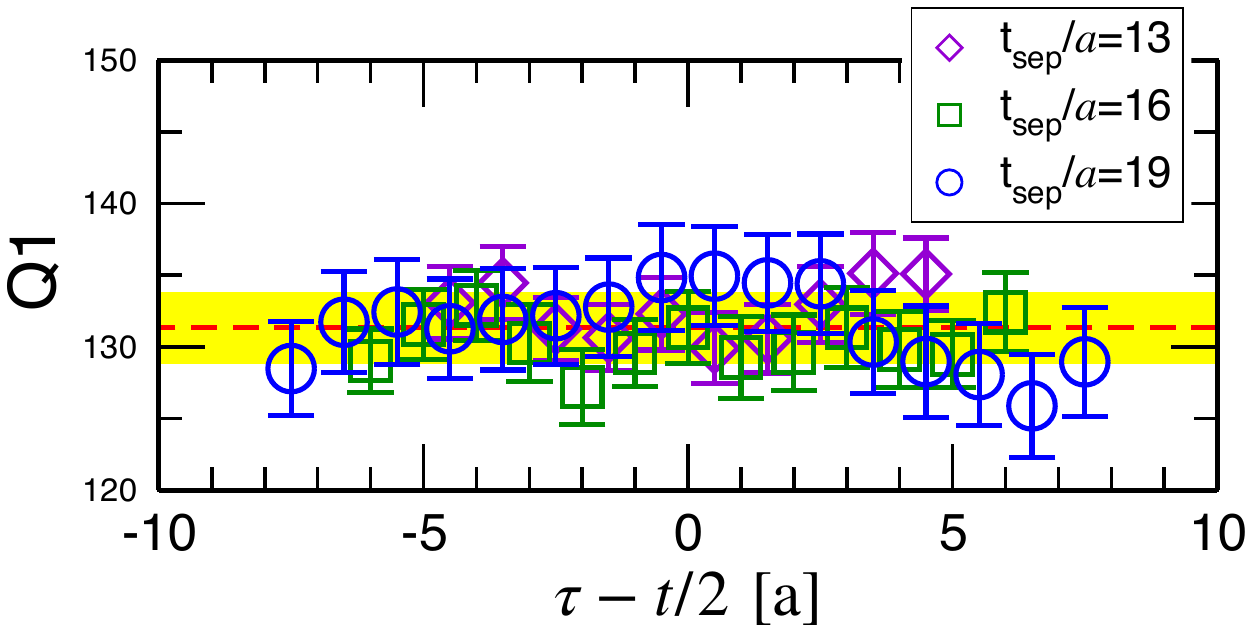}
\end{minipage}\hspace{0.03\linewidth}\begin{minipage}[tbp]{0.46\linewidth}\vspace{3pt}
\captionof{figure}{Results for $\tilde{G}_P(Q_1^2)$ obtained with the subtraction method developed in Ref.~\cite{Aoki:2025taf}. 
These very flat results should be compared to the top right plot in Fig.~\ref{fig:ratio_g4g5}. Within statistical errors, 
the data exhibit a plateau at different source-sink separations, indicating that the dominant $N\pi$ contribution has been removed. For more details, see Ref.~\cite{Aoki:2025taf}.
}
\label{fig:pacs_submethod}
\end{minipage}
\end{figure}

\paragraph{Variational analysis}
The most robust approach is a variational analysis. 
By constructing a sufficiently rich operator basis and solving the GEVP, one can isolate the relevant states and directly remove the dominant excited-state contributions from both two- and, in particular, three-point functions. The effectiveness of this method relies critically on including in the variational basis those operators that couple strongly to the current-enhanced states identified for the given channel and kinematics. Ref.~\cite{Barca:2025det} provides a systematic way to identify these most relevant operators. 
In Refs.~\cite{Barca:2022uhi, Barca:2024hrl}, it was sufficient to include only the current-enhanced operators.
In practice, constructing a large variational basis requires an efficient framework for computing the corresponding matrices of two- and three-point correlation functions.
The distillation method~\cite{HadronSpectrum:2009krc}, provides a systematic and flexible approach for constructing extended multi-hadron operators. See Refs.~\cite{PhysRevD.91.114501, Egerer:2021ymv} for some applications.
Distillation can also be combined with stochastic techniques to construct local current insertions efficiently, as introduced in Ref.~\cite{Hu:2025vhd}.
Beyond removing ESC, the variational framework enables direct access to transition matrix elements, such as $\langle N\pi|\mathcal{J}|N\rangle$~\cite{Barca:2024sub}, $\langle \pi\pi/\rho|\mathcal{V}_\mu| B\rangle$~\cite{Leskovec:2025gsw}, and excited-meson radiative transitions~\cite{PhysRevD.91.114501}.
\paragraph{Multi-state fits and related approaches}
Cheaper alternatives to a full variational analysis can also mitigate ESC.  
Since the volume-enhanced contributions originate from current-meson diagrams, the energy $E_M$ of the dominant mesonic state can be extracted from the current-meson two-point function,
$
\langle \mathrm{O}_M(\vec{q}, t)\,\mathcal{J}(\vec{q},\tau)\rangle 
\propto e^{-E_M(t-\tau)}
$.
This information can be used as a prior in multi-state fit ans\"atze by putting priors to $E_{HM}\simeq E_H+E_M$, or to subtract the dominant excited-state contribution directly~\cite{Aoki:2025taf}.
If a single excited state dominates, such as $N\pi$ in the nucleon pseudoscalar and axial channels, one can construct suitable linear combinations of three-point functions and their temporal derivatives to cancel this contribution~\cite{Aoki:2025taf,Tsuji:2025quu}, see Fig.~\ref{fig:pacs_submethod}. This approach is particularly effective in large volumes, where $N\pi$ states are approximately non-interacting, but may become less reliable in smaller volumes or when current-enhanced states are resonances.

Recently, Ref.~\cite{Wang:2025nsd} implemented the idea of Ref.~\cite{Barca:2025det} in a simplified variational setup. Instead of solving a full GEVP, they construct an optimal interpolator
$
\mathrm{O}_N^{\rm opt} = \mathrm{O}_N + c_{\rm opt}\,\mathrm{O}_{NM},
$
where $c_{\rm opt}$ needs to be determined, and the optimal operator is used to cancel the leading ESC in nucleon tensor and scalar charges, $g_T$ and $g_S$, respectively. In a full variational analysis, the coefficient $c_{\rm opt}$ corresponds to the nucleon component of the GEVP eigenvector. Using this method, they obtain the most precise lattice determinations of $g_T$ and $g_S$.
\vspace{-1em}
\paragraph{ChPT predictions for excited-state contribution}
Excited-state effects can also be estimated using ChPT.
Guided by the current-enhancement mechanism, one can identify the dominant intermediate states in each channel: 
$N\pi$ in nucleon axial and pseudoscalar three-point functions, consistent with Refs.~\cite{Bar:2018xyi, Bar:2019gfx, RQCD:2019jai}; 
$N(\vec{0})\pi(\vec{0})\pi(\vec{0})$ or $N\sigma$ in the isoscalar scalar nucleon channel~\cite{Gupta:2021ahb}; 
$N(\vec{0})\pi(\vec{1})\pi(-\vec{1})$ or $N\rho$ in the isovector vector nucleon channel~\cite{Barca:2025det, Green:2014xba};
$B^*K$ in $B_s \to K \ell \nu_\ell $ decays \cite{FermilabLattice:2019ikx}; $D^*\pi$ in $D\to \pi \ell \nu_\ell$ decays \cite{FermilabLattice:2022gku}; and many more.

Since the argument relies only on symmetry considerations and spatial-volume enhancement of specific diagrams, it is quite general and applicable to a wide class of hadronic three-point functions. 
For instance, it can be applied to the computation of nucleon PDFs and GPDs~\cite{Ji:2026vir, Gao:2026hix}.
 \vspace{-0.8em}
\section{Conclusions}
\vspace{-0.8em}
I have presented evidence that ESC in three-point functions can be enhanced and substantially larger than commonly assumed based on naive volume suppression arguments. Although multi-hadron overlap factors are suppressed by powers of the spatial volume, this suppression can be compensated by a corresponding volume enhancement in the current matrix elements for specific states. As a result, current-induced multi-hadron contributions can remain unsuppressed and significant at the source–sink separations accessible in present-day simulations.
In the nucleon sector, this effect is particularly pronounced in the pseudoscalar and axial channels, where the observed time dependence is consistent with dominant $N\pi$ intermediate states. Most importantly, if these contributions are not properly accounted for, the extracted form factors violate badly the generalised Goldberger-Treiman relation. 
Once the relevant multi-hadron states are included, the symmetry relation is restored. 
This demonstrates that uncontrolled ESC can render matrix elements unreliable. A systematic treatment of the relevant ESC is therefore a necessary condition for precision lattice QCD determinations. The mechanism discussed here is general and applies to hadronic three-point functions beyond the nucleon case. I hope that the arguments presented in this work, will provide practical guidance for future analyses aiming at reliable determinations of hadron matrix elements.
 \vspace{-0.7em}
\paragraph{Acknowledgements} 
I thank the Local Organising Committee for the invitation to present this plenary talk.  
I gratefully acknowledge illuminating discussions and collaboration with G. Bali and S. Collins, 
as well as valuable discussions with J.~Green, R.~Gupta, K.-F.~Liu, A.~Patella, S.~Prelovšek, S.~Schaefer, and R.~Sommer. 
I thank C.~Alexandrou and Y.~Li for sharing their data and further confirming 
the volume scaling of quark-line disconnected diagrams. This work was supported 
by the German Research Foundation (DFG) through the research unit FOR5269
“Future methods for studying confined gluons in QCD.”
\vspace{-1em}
 
\bibliographystyle{abbrvnat}
\bibliography{references}

\end{document}